\documentclass[12pt,preprint]{aastex}

\shorttitle{Major electron events and coronal magnetic
configurations} \shortauthors{Li et al.}

\begin{document}


\title{Major electron events and coronal magnetic configurations of the
related solar active regions}


\author{C. Li, C. J. Owen, and S. A. Matthews}
\affil{Mullard Space Science Laboratory, University College London,
Dorking, Surrey RH5 6NT, UK} \email{cl2@mssl.ucl.ac.uk}

\and

\author{Y. Dai, and Y. H. Tang}
\affil{Department of Astronomy, Nanjing University, Nanjing 210093,
PR China}




\begin{abstract}
A statistical survey of 26 major electron events during the period 2002
February through the end of solar cycle 23 is presented. We have
obtained electron solar onset times and the peak flux spectra for each
event by fitting to a power-law spectrum truncated by an exponential
high-energy tail, i.e., $f(E) \sim E^{-\delta}e^{-E/E_{0}}$. We also
derived the coronal magnetic configurations of the related solar active
regions (ARs) from the potential-field source-surface (PFSS) model. It
is found that: (1) 10 of the 11 well-connected open field-line events
are prompt events whose solar onset times coincide with the maxima
of flare emission. 13 of the 14 closed field-line events are delayed
events. (2) A not-well-connected open field-line event and one of the
closed field-line events are prompt events, they are both associated
with large-scale coronal disturbances or dimming. (3) An averaged harder
spectrum is found in open field-line events compared with the closed
ones. Specifically, the averaged spectral index $\delta$ is of 1.6 $\pm$
0.3 in open field-line events and of 2.0 $\pm$ 0.4 in closed ones. The
spectra of three closed field-line events show infinite rollover
energies $E_{0}$. These correlations clearly establish a significant
link between the coronal magnetic field-line topology and the escape of
charged particles from the flaring ARs into interplanetary space during
the major solar energetic particle (SEP) events.
\end{abstract}


\section{Introduction}

Non-thermal electrons are one population of particle streams
produced by the rapid release of magnetic energy during solar
eruptions, notably flares and coronal mass ejections (CMEs). In-situ
observations \citep{lin85,kru99} show that impulsive $\sim$ keV
electron events exhibit a rapid onset with the inferred release time
coincident with the soft X-ray (SXR), hard X-ray (HXR) emission, and
the type III radio burst. The relatively more gradual events
with energies up to MeV, namely major electron events, are sometimes
released several minutes later and accompanied by enhanced proton
fluxes.

It is generally agreed that major electron events are produced by
shock-wave acceleration \citep{rea99}. However, these events are
always associated with flares and CMEs, representing different
manifestations of the same magnetic energy release process
\citep{zha01}, both of which are capable of accelerating particles.
On the other hand, the coronal magnetic configurations are very
complex and could therefore provide tunnels for charged particles
escaping from the local coronal sites, i.e., flaring active regions
(ARs) into interplanetary space \citep{per86}. Recently, a number of
authors have compared open magnetic flux tubes with the solar
sources of impulsive SEP events \citep{wan06,nit06,rus08,mas09}.

The particle dynamics (for instance release times and spectra) of major
electron events show great complexity and differ from one event to
another. Apart from dependence due to acceleration mechanisms
\citep{kon01} and interplanetary transport effects \citep{can03}, one
viewpoint is that the coronal magnetic configuration might also
contribute to the dynamical discrepancies. In this letter, we present
a survey of 26 major electron events during the period 2002 February
through the end of solar cycle 23. Our purpose is to clarify the
correlation between the dynamics of in-situ electrons and the magnetic
field-line topologies in the vicinity of the related solar ARs.

\section{Observations}

Two spacecraft are currently orbiting the Sun-Earth L1 libration
point with experiments observing electrons from solar wind energies
up to the relativistic energy range. The WIND three-dimensional
Plasma and Energetic Particles instrument (3DP; Lin et al. 1995)
provides electron measurements with the electrostatic analyzers
(EESAs) from $\sim$ 0.5 keV to 28 keV, and the solid-state
telescopes (SSTs) from 27 keV to $\sim$ 300 keV. The Electron,
Proton, and Alpha Monitor (EPAM; Gold et al. 1998) onboard Advanced
Composition Explorer (ACE) measures electrons in the energy range of
38 -- 315 keV.

Our survey of these datasets began in 2002 February (start of the
RHESSI observations; Lin et al. 2002) and continued through the end
of solar cycle 23 (2006 December). The following selection criteria
were applied to identify the major electron events reported in this
study: (1) Electron intensity has an enhancement detected by ACE/EPAM,
which means that the in-situ electron energy is above $\sim$ 40 keV.
(2) The concurrent proton flux with energy above $\sim$ 10 MeV is
observed by Geostationary Operational Environment Satellite (GOES)
spacecraft. The proton event list is maintained by D. C. Wilkinson
(http://www.ngdc.noaa.gov/stp/GOES/goes.html). (3) There are obvious
distinctions between successive events, to ensure that any events
can be attributed to specific flare eruptions and the related solar
ARs. Additionally, the related solar ARs are located on the earthward
solar surface, to ensure that the magnetic field extrapolations can be
applied. These criteria reduced the candidates to 26 events listed in
Table 1.

\section{Data analysis}

The arrival times of near-relativistic electrons are recorded by
ACE/EPAM. Assuming electrons travel along the nominal Parker-spiral
interplanetary magnetic field (IMF) lines at a speed of $\upsilon$
with no scattering \citep{kru99}, and with respect to the flare
emission time, we estimate the SEP solar onset time by subtracting
$\Delta t=\ \rm \ell/\upsilon-8.3\,\rm$ minutes (Table 1, column 6)
from the in-situ onset time (column 5). The length of IMF lines
$\ell$ is calculated by solution of the IMF equation deduced from
the solar wind model \citep{par58}, $\upsilon$ is taken to be
$\sim$ 0.5$c$ in the energy channel of 53 -- 103 keV. As indicated
by \citet{hag03}, the in-situ onset in this channel leads to a lower
bound on the SEP solar onset time because of the residual straggling
effect (higher energy electrons leave a fraction of their energy in
the detector and are thus counted in the lower channels). This is
reasonable since electrons actually undergo interplanetary scattering
more or less. The inferred SEP solar onset times are then compared with
the peak times of HXR emission (RHESSI 50 -- 100 keV, h in column 4),
or if no HXR data available, with SXR emission (GOES 1 -- 8 $\rm{\AA}$,
s in column 4). A prompt event (P in column 7) is defined if the SEP
solar onset time is before or at the flare emission peak time. Otherwise,
the event is categorized as a delayed one (D in column 7).

The in-situ electron spectra generally show power-law or broken
power-law shapes (Lin et al. 1982). However, observations and
simulations show that the spectral profiles could somewhat deviate
from power-law distribution and display gradual changes at
high-energy tail due to a variety of effects
\citep{eli85,ana97,ham05,liu09}. Major SEP events are always
associated with both flares and CMEs. To avoid the argument of which
acceleration source dominates the production of charged particles
and which acceleration mechanism determines the shape of the
particle spectra, we applied a spectral profile with the combination
of a power-law and an exponential function, i.e., $f(E) \sim
E^{-\delta}e^{-E/E_{0}}$. Using the electron data from WIND/3DP
(EESAs and SSTs), the peak flux spectra for each event are then
fitted to the function, and the spectral indices $\delta$ and the
rollover energies $E_{0}$ are listed in Table 1, column 8 and 9,
respectively. The averaged value and standard deviation of the
spectral indices is $\delta=1.8 \pm 0.4$, which is similar to the
results of \citet{eli85}.

Figure 1 shows the comparison of a prompt event which occurred on
2002 August 14 with a delayed event on 2002 November 9. It is found
that the SEP solar onset time is consistent with the SXR, microwave
emission, and the type III radio burst in the prompt event. On the
contrary, electrons are released several minutes later compared to
the multi-wavelength flare emission in the delayed event. The
spectral index $\delta$ is of 1.4 in the prompt event much harder
than the index of 2.0 in the delayed event, and the rollover energy
$E_{0}$ in the prompt event is of 105 keV lower than the 330 keV in
the delayed event.

We then use the PFSS model developed by \citet{sch03}, which is
available in the IDL-based solar software (SSW) package, to identify
the coronal magnetic configurations. It has been successfully applied
in relating the large-scale topologies to the coronal plasma outflows
\citep{sak07} and to the open field-line fluxes \citep{li09}. Using
the photospheric longitudinal magnetogram from the Michelson Doppler
Imager (MDI) onboard the Solar and Heliospheric Observatory (SOHO),
the coronal magnetic configuration of each event is extrapolated.
The region of the extrapolation is in the vicinity of the solar AR,
the time of the selected magnetogram is just previous to the flare
eruption, and the height of the calculation is extending to the.
assumed solar source surface (at $\sim$ 2.5 $R_{\odot}$ where the
magnetic energy density equals to the plasma energy density)
Figure 2 shows the comparison of an open magnetic field-line topology
and a closed one, corresponding to the events plotted in Fig. 1. Of
these 26 major electron events, 12 are open field-line events (O in
column 10) and the remained 14 events are closed ones (C in column 10).

To further examine whether the open field-line fluxes are connected
to the IMF lines, which are rooted on the solar source surface and
linked to the near-Earth spacecraft, we calculate the connection
longitude as $\phi = \omega(r-r_{0})/u$. Where $r$ is the distance
from the Sun center to the L1 libration point, $r_{0}$ is the radius
of the solar source surface, $\omega$ is the angular speed of solar
rotation, and $u$ is the solar wind speed. Taking into account the
uncertainty of about $\pm 10^{\circ}$ \citep{ipp05}, the connection
longitudes are derived for each event. As shown in Fig. 2, the open
field-line fluxes of the 2002 August 14 event intersect the connection
longitudes, suggesting a so-called well-connected event. Of the 12
open field-line events, only the 2004 July 25 event is not
well-connected. The latitudinal connection is not considered in this
study, since as indicated by \citet{kle07} that the detection of
Langmuir waves with the kilometric type III bursts suggesting the IMF
lines bend down to the ecliptic from higher latitude solar source
surface.

\section{Results and discussion}

Based on the above analysis, significant correlations could be
established. Of the 11 well-connected open field-line events, 10
(91$\%$) are prompt events. The exceptional delayed event occurred
on 2003 November 2. Since the PFSS model does not correctly rebuild
the local and delicate magnetic structures such as twist and
helicity, a possibility is that the magnetic configurations in the
lower coronal site are not favorable for releasing charged particles.
Of the 14 closed field-line events, 13 (93$\%$) are delayed events.
The exceptional prompt event occurred on 2005 August 22. From the
based-difference 195 {\AA} images (see Fig. 3) of the
Extreme-Ultraviolet Imaging Telescope (EIT) on board SOHO, it is
found that large-scale coronal disturbances or dimming take place
around the SEP solar onset time (01:18 UT) as shown by the arrows. This
process might open or significantly reconfigure the magnetic field in
the corona, facilitating the escape of charged particles \citep{li06}.
We note that the not-well-connected open field-line event on 2004
July 25 has similar coronal signatures, explaining the prompt
injection of non-thermal electrons.

Furthermore, the averaged values and standard deviations of spectral
indices of the open field-line events and the closed ones are
$\delta_{open} =\ $1.6 $\pm$ 0.3 and $\delta_{closed} =\ $2.0 $\pm$
0.4, respectively. Additionally, the spectra of three closed
field-line events show infinite rollover energy $E_{0}$ (see Table 1,
column 9). The differences may result from the so-called ``mixed" or
``hybrid" acceleration processes \citep{kal03}. Charged particles in
the open field-line event can escape more easily from the flare
acceleration region into higher coronal regions and be re-accelerated
by the CME-driven shock, producing a more intense event and a harder
spectrum \citep{lin82}. A great number of charged particles in the
closed field-line event are supposed to be trapped in the flaring
ARs, providing fewer particles to the further shock acceleration,
thus the spectral shape could be steeper and follow a power-law
distribution extending to high energies. A larger sample size of major
or impulsive electron events should be further studied to reduce the
uncertainties on the spectral discrepancies.

In this Letter, significant correlations are established between the
coronal field-line topologies and the dynamics of in-situ electrons.
To conclude, in addition to being accelerated by CME-driven shocks
during the major SEP events, both the flare acceleration and the
coronal magnetic configuration could play important role in producing
and guiding charged particles from the low coronal site into
interplanetary space.

\acknowledgments

We are very grateful to the anonymous referee, whose constructive
comments greatly improved this work. It is a pleasure to thank
the WIND/3DP, ACE/EPAM, RHESSI, GOES, and SOHO teams for providing
the data used in this study. This work was supported by the rolling
grant from Science $\&$ Technology Facilities Council (STFC) of the
UK.

\begin{tiny}
\begin{deluxetable}{lccccccccc}

\addtolength{\tabcolsep}{-4pt}
\renewcommand{\arraystretch}{1}

\tablecolumns{10} \tablewidth{0pc} \tablecaption{Major electron
events, 2002 February - 2006 December} \tablehead{\colhead{} &
\colhead{} & \colhead{} & \colhead{Emission} & \colhead{In-situ} &
\colhead{}& \colhead{} & \colhead{} & \colhead{} & \colhead{}\\
\colhead{} & \colhead{Flare} & \colhead{Flare} & \colhead{Peak} &
\colhead{Onset} & \colhead{$\Delta t$} & \colhead{SEP} & \colhead{} &
\colhead{$E_{0}$} & \colhead{PFSS}\\
\colhead{Date} & \colhead{Location} & \colhead{Class} &
\colhead{(UT)}  & \colhead{(UT)} & \colhead{(min)} & \colhead{Type} &
\colhead{$\delta$} & \colhead{(keV)}  & \colhead{Model}}

\startdata
2002 Feb 20 & N12W72& M5.1& 06:12 s& 06:06& 10.4 & P & 1.5& 140& O \\
2002 Mar 16 & S08W03& M2.2& Mar 15 23:10 s& 00:57& 11.0 & D & 1.8& 345& C \\
2002 Mar 18 & S09W46& C5.9& 11:45 s& 12:33& 10.7 & D & 2.1& $\infty$& C \\
2002 Apr 17 & S14W34& M2.9& 08:57 s& 08:43& 10.9 & P & 1.8& 100& O \\
2002 Apr 21 & S14W84& X1.5& 01:46 h& 01:38& 10.1 & P & 1.1& 346& O \\
2002 May 22 & S19W56& M1.0& 00:20 s& 00:26& 10.6 & P & 2.2& 114& O \\
2002 Aug 14 & N09W54& M2.3& 02:12 s& 01:59& 10.2 & P & 1.4& 105& O \\
2002 Aug 22 & S07W62& M5.4& 01:52 h& 02:18& 10.4 & D & 2.1& 249& C \\
2002 Nov 9  & S12W29& M4.6& 13:23 h& 14:02& 10.8 & D & 2.0& 212& C \\
2003 May 28 & S07W17& X3.6& 00:52 h& 00:56& 9.6  & P & 1.5& 860& O \\
2003 May 31 & S07W65& M9.3& 02:29 h& 02:40& 9.4  & D & 1.6& 112& C \\
2003 Oct 26 & N02W38& X1.2& 18:19 s& 17:52& 10.2 & P & 1.7& 537& O \\
2003 Oct 28 & S16E08& X17.2& 11:06 h& 11:20& 9.5 & D & 2.7& 100& C \\
2003 Nov $2^{a}$ & S14W56& X8.3& 17:17 h& 17:40& 9.9 & D & 1.5& 309& O \\
2004 Apr 11 & S14W47& C9.6& 04:19 s& 04:27& 10.3 & P & 2.1& 199& O \\
2004 Jul 25 & N08W33& M1.1& 15:14 s& 15:21& 9.8  & P & 1.7& 980& $\rm O^{N}$ \\
2004 Sep 19 & N03W58& M1.9& 17:12 s& 17:31& 10.7 & D & 2.0& $\infty$& C \\
2004 Nov 7  & N09W17& X2.0& 16:20 h& 17:26& 10.1 & D & 2.1& $\infty$& C \\
2005 Jan 15 & N15W05& M9.1& 06:35 s& 07:10& 9.8  & D & 2.5& 528& C \\
2005 Jan 20 & N14W61& M7.1& 06:44 h& 06:54& 9.3  & D & 1.5& 474& C \\
2005 May 13 & N12E11& M8.0& 16:50 h& 17:29& 9.7 & D & 2.2& 259& C \\
2005 Jun 16 & N09W87& M4.0& 20:10 h& 20:35& 9.5 & D & 1.7& 324& C \\
2005 Jul 13 & N10W80& M5.0& 14:49 s& 14:37& 9.7 & P & 1.0& 76& O \\
2005 Aug $22^{b}$ & S10W52& M2.8& 01:17 h& 01:19& 9.9 & P & 1.6& 186& C \\
2006 Dec 5 & S07E79& X9.0& 10:30 h& 14:07& 11.3 & D & \nodata& \nodata& C \\
2006 Dec 13 & S05W23& X3.4& 02:40 s& 02:48& 9.5 & P & \nodata& \nodata& O \\
\enddata

\tablecomments{In Column 4, s indicates that the emission peak time
is from SXR observation, and h from HXR. In Column 7, P indicates a
prompt SEP event, and D is delayed. In Column 10, O indicates a open
field-line event, and C is closed. Dotted line indicates no
data available.}

\tablenotetext{a}{The 2003 November 2 event is a delayed event,
however, corresponds to an open field-line topology. $^{b}$The 2005
August 22 event is a prompt event, however, corresponds to a closed
field-line topology.}

\tablenotetext{N}{Open field-line fluxes are not well-connected to the
IMF lines connecting the solar source surface to the near-Earth spacecraft.}

\end{deluxetable}
\end{tiny}

\begin{figure}
\epsscale{.80} \plotone{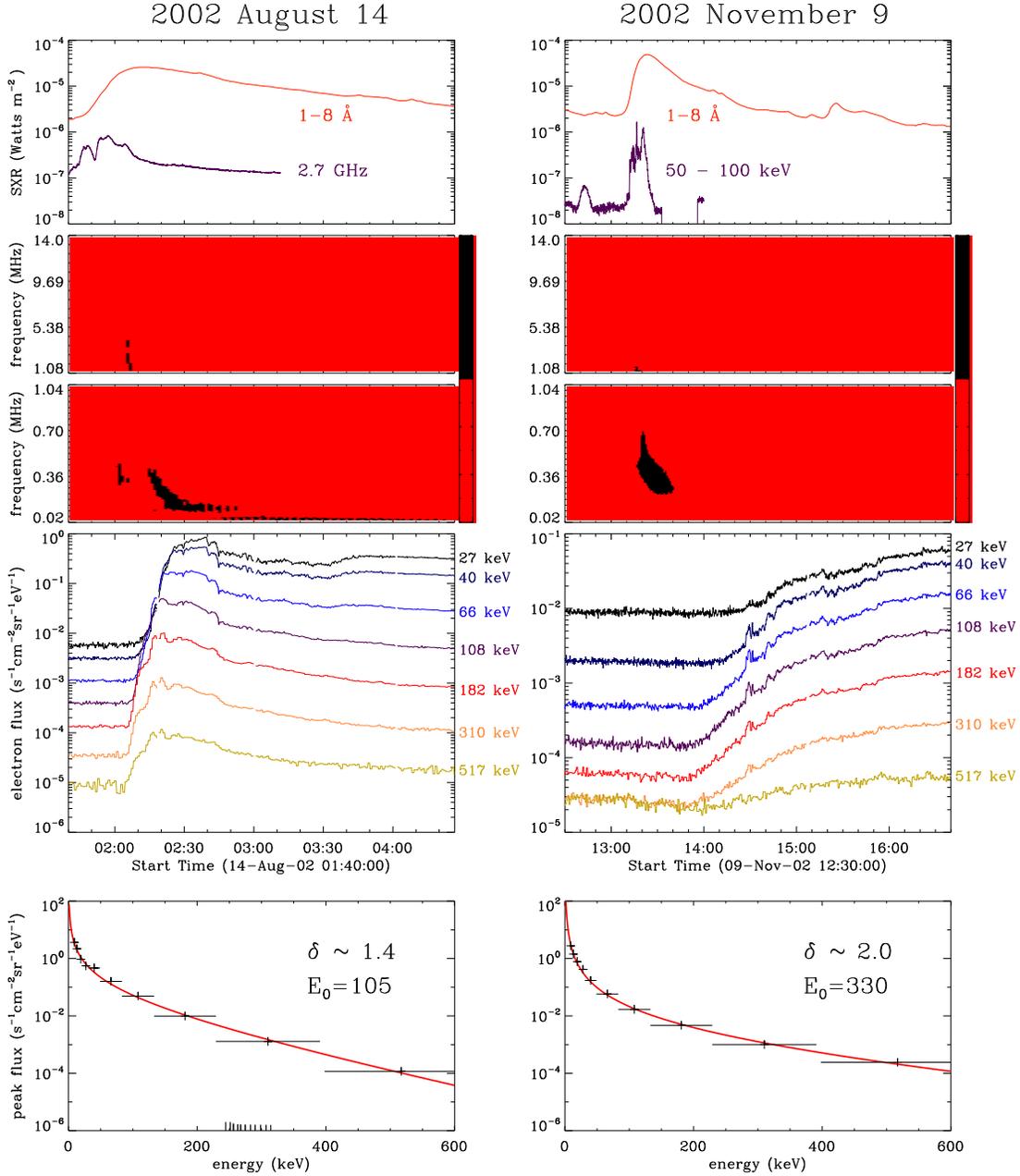} \caption{Prompt and delayed
electron event observed on 2002 August 14 and 2002 November 9,
respectively. From top to bottom: GOES SXR 1 -- 8 $\rm{\AA}$,
RSTN/Learmonth microwave 2.7 GHz (left), and RHESSI HXR 50 -- 100
keV light curves (right); WIND/WAVES radio spectrograms at frequency
range of 20 kHz -- 14 MHz; WIND/3DP in-situ observed electrons from
$\sim$ 30 keV to $\sim$ 500 keV; Electron peak flux spectra in a
linear-log scale. \label{fig1}}
\end{figure}

\begin{figure}
\epsscale{.80} \plotone{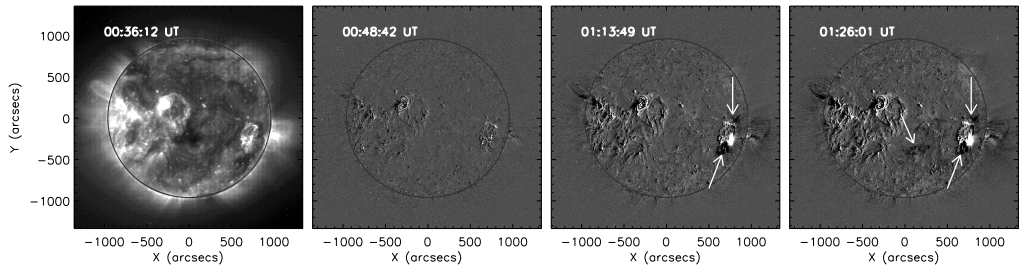} \caption{The coronal disturbance or
dimming observed on 2005 August 22. The pre-event SOHO/EIT 195
$\rm{\AA}$ image is shown in the left, and the following three are
difference images with the pre-event image subtracted from them. Arrows
show the regions of disturbance or dimming takes place.
\label{fig3}}
\end{figure}

\end{document}